\documentclass{article}
\usepackage{graphics} 
\begin{document}
\begin{center}
EFFECTIVE ACTION AND MEAN FERMION NUMBER DENSITY OF GRAPHENE IN CONSTANT MAGNETIC FIELD AT 
FINITE TEMPERATURE AND DENSITY

\vspace{0.5cm}

Alok Kumar{\footnote{e-mail: alok.academic@gmail.com {\it{[alok[POINT]academic[AT]gmail[POINT]com]}}}} \\
Harish-Chandra Research Institute\\
Chhatnag Road, Jhusi\\
Allahabad 211 019\\
Uttar Pradesh,India.\\
\end{center}

\vspace{1.0cm}

{\noindent{\it{Abstract}}} 

\vspace{0.5cm}

The effective action and the mean fermion number density of graphene in constant external 
magnetic field at finite temperature and density are calculated. Closed expressions for 
these are given and their variation with temperature are studied. It is found that the 
mean fermion number density peaks around a particular temperature, depending on the chemical 
potential at low temperatures. This feature is interpreted as 'condensation of 
fermions $\bar{\psi}\psi$' in graphene. In future, it is interesting to extend and explore
 this calculation and the work of the reference [20] for the case of {\bf{Graphyne}} [23].

\vspace{1.5cm}

Graphene is a honeycomb lattice of Carbon atoms with remarkable electronic properties [1,2].
Many of these properties are a low energy phenomena, governed by massless Dirac equation for the 
electrons in the tight-binding (hopping) model. This has been discovered in a seminal 
contribution of Semenoff [3] and exemplified by Mecklenberg and Regan [4] and Ando [5]. Thus, 
the excitations of graphene in the tight-binding approximation near the Fermi surface are 
described by (2+1)-dimensional massless Dirac equation , with velocities $v_F\sim 10^6\ ms^{-1}$.
This description is accurate theoretically [3], [6], neglecting many body effects and 
experimentally [7,8], as the quasiparticles in graphene exhibit {\it{the linear dispersion 
relation}} $E=pv_F$ (as if they are massless relativistic particles of speed $v_F$) and 
by the observation of a relativistic analogue of integer quantum Hall effect. {\it{Recently, it is shown mathematically in [20], 
that the linear dispersion relation (Dirac points) is generic in (non-relativistic) quantum crystals having the Honeycomb lattice symmetry.}} It is 
important to realize that the quasi-particles of graphene have to be described by two-component
wave functions, necessary to define relative contributions of sublattices [9]. Such a 
two-component description is similar to spinor wave functions in QED, with the 'spin index' 
for graphene indicating sublattices rather than the real spin of electrons. 

\vspace{0.5cm}

A graphene layer has been extracted out of graphite [10] and such a discovery of 2-dimensional 
system attracted tremendous interest in theoretical physics, such as massless Dirac fermion, 
quantum Hall effect and even fractional quantum Hall effect [11], providing a framework to use 
the methods developed in quantum gauge field theories in graphene. Many of  graphene physics is described by the Quantum Field Theory (QFT) [21].
{\it{It is remarkable to note that many High Energy Physics (HEP) Experiments can be realised with the graphene table-top experiments. This statements is written in the spirit of Moore's Law for Electronics.}} It is proposed in [22] how certain experimental results for carbon nanotubes (very closely related to graphene) could be re-interpreted in terms of the Kaluza-Klein mode of extra dimension. {\it{Now the physics is more interesting with the advent of the new material {\bf{Graphyne}} [23], which is more versatile than graphene due to its direction-dependent dirac cones.}}
A functional determinant approach with zeta function regularization has been employed by Beneventano and Santangelo [12] 
to study the quantum Hall effect in graphene and by Beneventano, Giacconi, Santangelo and Soldatic
[13] to study the one-loop effects for massless Dirac fields coupled to electromagnetic 
backgrounds both at zero and finite temperature and density. The present author have studied 
$SU(2)$ Yang-Mills theory at finite temperature in (3+1)-dimensions for understanding the 
vacuum in Quantum Chromo Dynamics [14] and so it is worthwhile to use the methods developed 
in [14] in graphene. 

\vspace{0.5cm}

It is the purpose of this contribution to derive closed expression for the effective action 
${\Gamma}_{eff}$ for graphene (quasiparticles in graphene) in an external constant magnetic 
field at finite temperature and density. The 'mean fermionic number density' $N_F=\frac{1}
{\beta}\ \frac{\partial {\Gamma}_{eff}}{\partial \mu}$, where $\mu$ is the chemical potential,
is calculated. We have used the $\epsilon$-regularization method of Salam and Strathdee [15] to 
regularize the divergences to obtain a finite action. The behaviour of the effective action and 
$N_F$ with temperature has not been reported so far in the literature to the best of our 
knowledge. 

\vspace{0.5cm}

Our results for ${\Gamma}_{eff}$ and $N_F$ are plotted against temperature for representative 
values of $\mu$. We find that the effective action peaks around a particular temperature for a 
given $\mu$. The 'mean fermionic number density' also peaks around the same temperature. This 
peaking of $N_F$ is interpreted as a possible 'fermion pair condensation' or 'quasi-particle 
pair condensation' in graphene.  

\vspace{1.0cm}

We consider massless electron in graphene coupled to an external magnetic field ($U(1)$ gauge 
field) in (2+1)-dimensions, described by the Lagrangian density
\begin{eqnarray}
{\cal{L}}&=&\bar{\psi}\ i{\gamma}^{\mu}D_{\mu}\ \psi; \ \ \vec{A}=(-i\frac{\mu}{e},0,Bx), 
\end{eqnarray}
where we use $\hbar=v_F=1$ units and the constant magnetic field is perpendicular to the plane of 
graphene. $\mu$ is the chemical potential for the electrons. For the  Euclidean version of (1) 
(useful to study the theory at finite temperature) we use  
\begin{eqnarray}
&{\gamma}^0=-{\sigma}_3\ ;\ {\gamma}^1={\sigma}_2\ ;\ {\gamma}^2={\sigma}_1\ ;\ \{ {\gamma}^i,
{\gamma}^j\}=2{\delta}^{ij},& \nonumber \\
&\slash\!\!\!\!D =i\slash\!\!\!{\partial}-e\slash\!\!\!\!A.&  
\end{eqnarray} 
The eigenvalue equation 
\begin{eqnarray}
(i\slash\!\!\!{\partial}-e\slash\!\!\!\!A)\psi &=&\lambda \psi,
\end{eqnarray}
gives 
\begin{eqnarray}
\lambda &=&\pm\sqrt{2NeB+({\omega}_n+\mu)^2},
\end{eqnarray}
where the harmonic oscillator quantum number $N=0,1,2,\cdots \infty$ and the Matsubara frequency 
${\omega}_n=(2n+1)\frac{\pi}{\beta};\ n=-\infty$ to $+\infty$, with $\beta=\frac{1}{kT}$, $k$ 
being the Boltzmann constant. Equation (4) gives the Landau levels of the fermions in graphene. 
From the (Euclidean) partition function 
\begin{eqnarray}
Z&=&\int [d\psi][d\bar{\psi}]\ e^{-\int d^3x\ {\cal{L}}_E}, \nonumber 
\end{eqnarray}
the effective action ${\Gamma}_{eff}=\ell og\ Z=\ell og\ det(i\slash\!\!\!{\partial}-e\slash 
\!\!\!\!A)$ is given by 
\begin{eqnarray}
{\Gamma}_{eff}&=&-\frac{1}{4\beta}\ \Big(\frac{eB}{2\pi}\Big)\ \sum_{N=0}^{\infty}\ \sum_{
n=-\infty}^{+\infty}\ \ell og\Big(\frac{1}{ {\Lambda}^2}\{2NeB+({\omega}_n+\mu)^2\}\Big), 
\end{eqnarray}
where the prefactor $\Big(\frac{eB}{2\pi}\Big)$ is the harmonic oscillator degeneracy factor and 
a dimensionful parameter $\Lambda$ is introduced to render the argument of the logarithm 
dimensionless. In odd dimensions (such as here), there is another inequivalent representation 
for the gamma matrices, obtained for example, by reversing the sign of one or all the three 
gamma matrices in (2). Inclusion of this introduces a factor two for ${\Gamma}_{eff}$. 

\vspace{0.5cm}

Now we use the $\epsilon$-regularization of Salam and Strathdee [15] (see also [16]) to write 
\begin{eqnarray}
{\Gamma}_{eff}&=&-\frac{1}{4\beta}\Big(\frac{eB}{2\pi}\Big)\sum_{N=0}^{\infty}\sum_{n=-\infty}
^{\infty}\Big(\frac{-i^{\epsilon}}{\epsilon\Gamma(\epsilon)}\Big)\ \int_0^{\infty} dt\ 
t^{-1+\epsilon} e^{-\frac{it}{ {\Lambda}^2}(2NeB+({\omega}_n+\mu)^2)},
\end{eqnarray}
in the limit $\epsilon\rightarrow 0$. After a Wick rotation $t\rightarrow -i\tau$ and carrying 
out the $N$-sum, we find 
\begin{eqnarray}
{\Gamma}_{eff}&=&-\frac{1}{4\beta}\Big(\frac{eB}{2\pi}\Big)\sum_{n=-\infty}^{\infty}
\Big(\frac{-1}{\epsilon\Gamma(\epsilon)}\Big)\int_0^{\infty}\frac{d\tau}{(1-e^{-2\tau eB/
{\Lambda}^2})} \nonumber \\
& & {\tau}^{-1+\epsilon}\ e^{-\frac{\tau}{ {\Lambda}^2}({\omega}_n+\mu)^2}. 
\end{eqnarray}
The $n$-sum can be done using Jacobi ${\theta}_3$-function to give 
\begin{eqnarray}
{\Gamma}_{eff}&=&-\frac{1}{4\beta}\Big(\frac{eB}{2\pi}\Big)
\Big(\frac{-1}{\epsilon\Gamma(\epsilon)}\Big)\int_0^{\infty}\frac{d\tau}{(1-e^{-2\tau eB/
{\Lambda}^2})}\ \ {\tau}^{-1+\epsilon} \nonumber \\
& & e^{-\frac{\tau}{ {\Lambda}^2}\{\frac{ {\pi}^2}{ {\beta}^2}+{\mu}^2+\frac{2\mu\pi}{\beta}\}}
\ {\theta}_3\Big(\frac{2\pi i\tau}{ {\beta}^2{\Lambda}^2}+\frac{2\mu i\tau}{\beta {\Lambda}^2}
,\frac{4\pi i\tau}{ {\beta}^2{\Lambda}^2}\Big).
\end{eqnarray} 
Using the property of ${\theta}_3$-function [17], namely 
\begin{eqnarray}
{\theta}_3(z,it)&=&t^{-\frac{1}{2}}\ e^{-\frac{\pi z^2}{t}}\ {\theta}_3(\frac{z}{it},
\frac{i}{t}), 
\end{eqnarray}
we find 
\begin{eqnarray}
{\Gamma}_{eff}&=&\frac{\Lambda}{8\sqrt{\pi}}\Big(\frac{eB}{2\pi}\Big)\Big(\frac{1}{\epsilon 
\Gamma(\epsilon)}\Big)\int_0^{\infty}d\tau\ \frac{ {\tau}^{-\frac{3}{2}+\epsilon}}{(1-
e^{-2\tau eB/{\Lambda}^2})} \nonumber \\ 
& & {\theta}_3(\frac{1}{2}+\frac{\mu\beta}{2\pi},\frac{i{\beta}^2
{\Lambda}^2}{4\pi \tau}).
\end{eqnarray}
Rewriting ${\theta}_3(z,t)$ as 
\begin{eqnarray}
{\theta}_3(z,t)&=&\sum_{\ell =-\infty}^{\infty}\ e^{i\pi t{\ell}^2}\ e^{2\pi iz\ell},
\end{eqnarray}
equation (10) becomes 
\begin{eqnarray}
{\Gamma}_{eff}&=&\frac{\Lambda}{8\sqrt{\pi}}\Big(\frac{eB}{2\pi}\Big)\ \Big(\frac{1}{\epsilon 
\Gamma(\epsilon)}\Big)\ \int_{0}^{\infty}\ d\tau\ \frac{ {\tau}^{-\frac{3}{2}+\epsilon}}
{(1-e^{-2\tau eB/{\Lambda}^2})} \nonumber \\
& &\{ 1+2\sum_{\ell =1}^{\infty}\ e^{-\frac{ {\beta}^2{\Lambda}^2{\ell}^2}{4\tau}}\ \cos(\ell \pi
+\ell \mu\beta)\}.
\end{eqnarray}

\vspace{0.5cm}

In (12), the term without the $\ell$-sum has the integral over $\tau$ is evaluated [18] and the 
limit $\epsilon\rightarrow 0$ is taken to give 
\begin{eqnarray}
{\Gamma}_{eff}^{(1)}&=&\frac{1}{4}\Big(\frac{eB}{2\pi}\Big)\ \sqrt{2eB}\ \zeta(-\frac{1}{2}), 
\end{eqnarray} 
where the superscript denotes the contribution to ${\Gamma}_{eff}$ coming from the term 
without the $\ell$-sum. This contribution corresponds to the effective action at zero 
temperature and zero chemical potential, and agrees with [13] and [19]. In (12), the term 
involving the $\ell$-sum is evaluated, first by writing 
\begin{eqnarray}
\frac{1}{(1-e^{-2eB\tau/{\Lambda}^2})}&=&\sum_{n=0}^{\infty}\ e^{-\frac{2neB}{ {\Lambda}^2}
\tau}\ =\ 1+\sum_{1}^{\infty}\ e^{-\frac{2neB}{ {\Lambda}^2}\ \tau}, \nonumber 
\end{eqnarray}
and then using the function $K_{\nu}(xz)$ [18] and taking the limit $\epsilon \rightarrow 0$, 
to give 
\begin{eqnarray}
{\Gamma}_{eff}^{(2)}&=&\frac{1}{4\pi}\Big(\frac{eB}{2\pi}\Big)\ \sum_{\ell=1}^{\infty}\ \cos(
\ell \pi+\ell \mu\beta)\{ \frac{2}{\ell \beta}\Gamma(\frac{1}{2}) \nonumber \\
&+&\sum_{n=1}^{\infty}\ \frac{2\sqrt{2}(2neB)^{\frac{1}{4}}}{\sqrt{\ell \beta}}\ K_{\frac{1}{2}}
(\ell \beta\sqrt{2neB})\},
\end{eqnarray}
so that the effective becomes (adding (13) and (14))
\begin{eqnarray}
{\Gamma}_{eff}&=&\frac{1}{4}\Big(\frac{eB}{2\pi}\Big)\sqrt{2eB}\ \zeta(-\frac{1}{2}) \nonumber \\
&+&\frac{1}{4\pi}\Big(\frac{eB}{2\pi}\Big)\ \sum_{\ell=1}^{\infty}\ \frac{\cos(\ell \pi+\ell \mu 
\beta)}{\beta\ell}\ \{2\sqrt{\pi} \nonumber \\
&+&\sum_{n=1}^{\infty}2\sqrt{2}\sqrt{\beta\ell\sqrt{2neB}}
K_{\frac{1}{2}}(\beta\ell\sqrt{2neB})\}.
\end{eqnarray}
Finally, using $K_{\frac{1}{2}}(z)=\frac{\sqrt{\pi}}{\sqrt{2z}}\ e^{-z}$ [18], the effective 
action (15) is written as 
\begin{eqnarray}
{\Gamma}_{eff}&=&\Big(\frac{eB}{2\pi}\Big)\ \Big(\frac{1}{4}\sqrt{2eB}\zeta(-\frac{1}{2}) 
\nonumber \\
&+&\frac{1}{2\sqrt{\pi}}\sum_{\ell=1}^{\infty}\frac{\cos(\ell\pi+\ell\mu\beta)}{\ell\beta}\{1+
\sum_{n=1}^{\infty}e^{-\beta\ell \sqrt{2neB}}\}\Big).
\end{eqnarray}
This is one of our main results, giving the effective energy density of the quasi-particles in 
graphene in a constant magnetic field at finite temperature and chemical potential. It is 
important to notice that the dimensionful parameter $\Lambda$, introduced in (5), has 
disappeared, unlike in (3+1)-dimensional gauge field theories [14,16]. The 'mean fermion number 
density' can be found using $N_F=\frac{1}{\beta}\frac{\partial {\Gamma}_{eff}}{\partial \mu}$.

\vspace{0.5cm}

In order to give the variation of ${\Gamma}_{eff}$ with temperature, it is convenient to introduce
\begin{eqnarray}
&x=\beta\sqrt{2eB};\ y=\frac{\mu}{\sqrt{2eB}};\ \mu\beta=xy,& 
\end{eqnarray}
so that 
\begin{eqnarray}
\frac{ {\Gamma}_{eff}}{\Big(\frac{eB}{2\pi}\Big)\sqrt{2eB}}&=&\frac{1}{4}\zeta(-\frac{1}{2})+
\frac{1}{2\sqrt{\pi}}\sum_{\ell=1}^{\infty}\ \frac{\cos(\ell\pi+\ell xy)}{\ell x}\{1+\sum_{n=1}
^{\infty}e^{-\ell\sqrt{n}x}\}, 
\end{eqnarray}
and 
\begin{eqnarray}
\frac{N_F}{\Big(\frac{eB}{2\pi}\Big)\sqrt{2eB}}&=&-\frac{1}{2\sqrt{\pi}}\sum_{\ell=1}^{\infty}
\frac{\sin(\ell\pi+\ell xy)}{x}\{1+\sum_{n=1}^{\infty}\ e^{-\ell\sqrt{n}x}\}, \nonumber \\
&=&-\frac{1}{2\sqrt{\pi}}\{-\frac{1}{2x}\tan\Big(\frac{xy}{2}\Big) \nonumber \\
&+&\sum_{\ell=1}^{\infty}\ 
\sum_{n=1}^{\infty}\ \frac{\sin(\ell\pi+\ell xy)}{x}\ e^{-\ell\sqrt{n}x}\}.
\end{eqnarray}

\vspace{0.5cm}

In Figure 1, $\frac{ {\Gamma}_{eff}}{\Big(\frac{eB}{2\pi}\Big)\sqrt{2eB}}$ is plotted against
$\frac{1}{x}=\frac{k}{\sqrt{2eB}}\ T$ for $y=0.5\ (\mu=0.5\sqrt{2eB})$. The effective energy 
density peaks around $\frac{1}{x}=0.18$. 
In Figure 2, $\frac{|N_F|}{\Big(\frac{eB}{2\pi}\Big)\sqrt{2eB}}$ is shown against $x^{-1}$ 
for $y=0.5$. The 'mean fermion number density' peaks around $x^{-1}=0.18 $. In Figures 3 and 4, 
we give them for $y=0.8$. We have found for $y=0.4, 0.5, 0.8, 1.0$, the peaking occurs at different 
values of $x^{-1}$. The minimum value of $y$ for the peak to occur is found to be $y=0.32$. 

\vspace{0.5cm}

This feature of peaking of $N_F$ is interpreted as 'condensation of quasi-particle charge 
density'. Since the fermion number or charge density in this framework is proportional to 
$\bar{\psi}\psi$, it is the pair $\bar{\psi}\psi$ (like Cooper pair) that condenses at low 
temperature and hence no violation of Pauli principle. This feature is different from the 
chiral symmetry breaking as it occurs around a particular temperature. Second, chiral symmetry 
breaking occurs when $\frac{\partial {\Gamma}_{eff}}{\partial m}|_{m=0}\neq 0$, where $m$ is 
the fermion mass. Since the role of $\mu$ is similar to that of $m$, we see from (18) and (19) that 
$\frac{\partial {\Gamma}_{eff}}{\partial \mu}|_{\mu=0}=0$ and so we realize a genuine 
condensation phenomenon. For the further deeper interpretation of the condensation we can look [24] and try to link our results in terms of material sciences. 

\vspace{1.5cm}

{\noindent{\bf{Acknowledgements}}}

\vspace{0.5cm}

We acknowledge useful discussions with Professor R. Parthasarathy (CMI, Chennai) and for helping in many ways, too. We acknowledge with 
thanks the Director of Chennai Mathematical Institute (CMI) for warm hospitality. We acknowledge  nicely - {\it{the XIth plan Neutrinos Physics Project, HRI, Allahabad}} - for supporting as a visitor during which the final version of this work came. 

\vspace{1.0cm}

{\noindent{\bf{References}}}

\vspace{0.5cm}

\begin{enumerate}
\item A.C.Neto, F.Guinea, N.M.R.Peres, K.S.Novoselov and A.K.Geim, Rev.Mod.Phys. {\bf{81}} 
(2009) 109. 
\item S.D.Sarma, S.Adam, E.H.Huang and E.Rossi, arXiv:1003.4731.
\item G.Semenoff, Phys.Rev.Lett. {\bf{53}} (1984) 2449.
\item M.Mecklenberg and B.C.Regan, arXiv:1003.3715. 
\item T.Ando, J.Phys.Soc.Japan, {\bf{74}} (2005) 777. 
\item F.D.M.Haldane, Phys.Rev.Lett. {\bf{61}} (1988) 2015.
\item K.S.Novoselov. et.al., Nature {\bf{438}} (2005) 438. 
\item Y.Zhang, Y.W.Tan, H.L.Stormer and P.Kim, Nature {\bf{438}} (2005) 201. 
\item M.I.Katsnelson, K.S.Novoselov and A.K.Geim, Nature Physics {\bf{2}} (2006) 620.
\item K.S.Novoselov, A.K.Geim, S.V.Morozov, D.Jiang, Y.Zhang, S.V.Dubonos, I.V.Grigorieva and 
      A.A.Firsov, Science {\bf{306}} (2004) 666. 
\item A.Kumar, J.M.Poumirol, W.Escoffier, M.Goiran, B.Raquet and J.C.Pivin, arXiv:1006.1085.    
\item C.G.Beneventano and E.M.Santangelo, arXiv:0710.4928: J.Phys. {\bf{A41}} (2008) 164035.
\item C.G.Beneventano, P.Goicconi, E.M.Santangelo, and R.Soldati, hep-th/0901.0396: J.Phys.
      {\bf{A42}} (2009) 275401.
\item R.Parthasarathy and Alok Kumar, Phys.Rev. {\bf{D75}} (2007) 085007. 
\item A.Salam and J.Strathdee, Nucl.Phys. {\bf{B90}} (1975) 203. 
\item D.Kay, A.Kumar and R.Parthasarathy, Mod.Phys.Lett. {\bf{A20}} (2005) 1655.
\item Bateman Manuscript (McGraw-Hill, New York, 1953), Vol.II.
\item I.S.Gradshteyn and I.M.Ryzhik, {\it{Table of Integrals, Series and Products}} (Academic 
      Press, New York, 1965). 
\item C.G.Beneventano, P.Giacconi, E.M.Santangelo and R.Soldati, hep-th/0701095: J.Phys.
      {\bf{A40}} (2007) F435.       
\item C.L. Fefferman and Michael I. Weinstein, Honeycomb Lattice Potentials and Dirac Points, arXiv:1202.3839v1 [math-ph].
\item I.V. Fialkovsky and D.V. Vassilevich, Quantum Field Theory in Graphene, arXiv:1111.3017v2[hep-th].
\item Jonas de Woul, Alexander Merle, and Tommy Ohlsson, Probing the Concept of Extra Dimensions with Carbon Nanotubes, arXiv:1203.3196v1 [hep-ph].
\item Daniel Malko, Christian Neiss, Francesc Vines, and Andreas Gorling, Competition for Graphene : Graphynes with Direction-Dependent Dirac Cones, PRL {\bf{108}},086804 (2012).
\item Oleg L. Berman, Roman Ya. Kezerashvili, and Yurii E. Lozovik, Bose-Einstein condensation of quasiparticles in graphene, arXiv:0908.3039v1 [cond-mat.mes-hall].
\end{enumerate} 

\newpage 
\begin{figure}
\begin{center}
\resizebox{120mm}{!}{\includegraphics{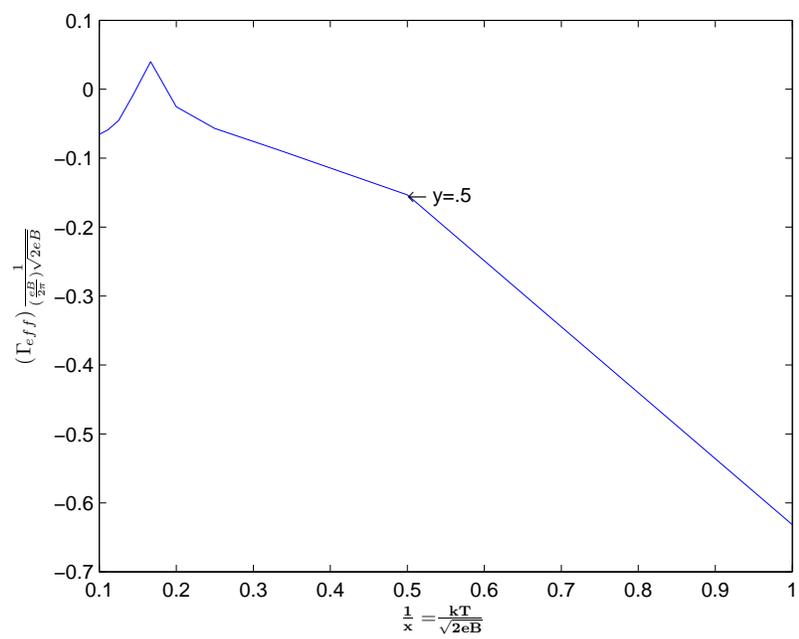}}
\caption{Variation of scaled effective energy density with scaled temperature for y=0.5}
\end{center}
\end{figure} 

\vspace{0.5cm}

\begin{figure}
\begin{center}
\resizebox{120mm}{!}{\includegraphics{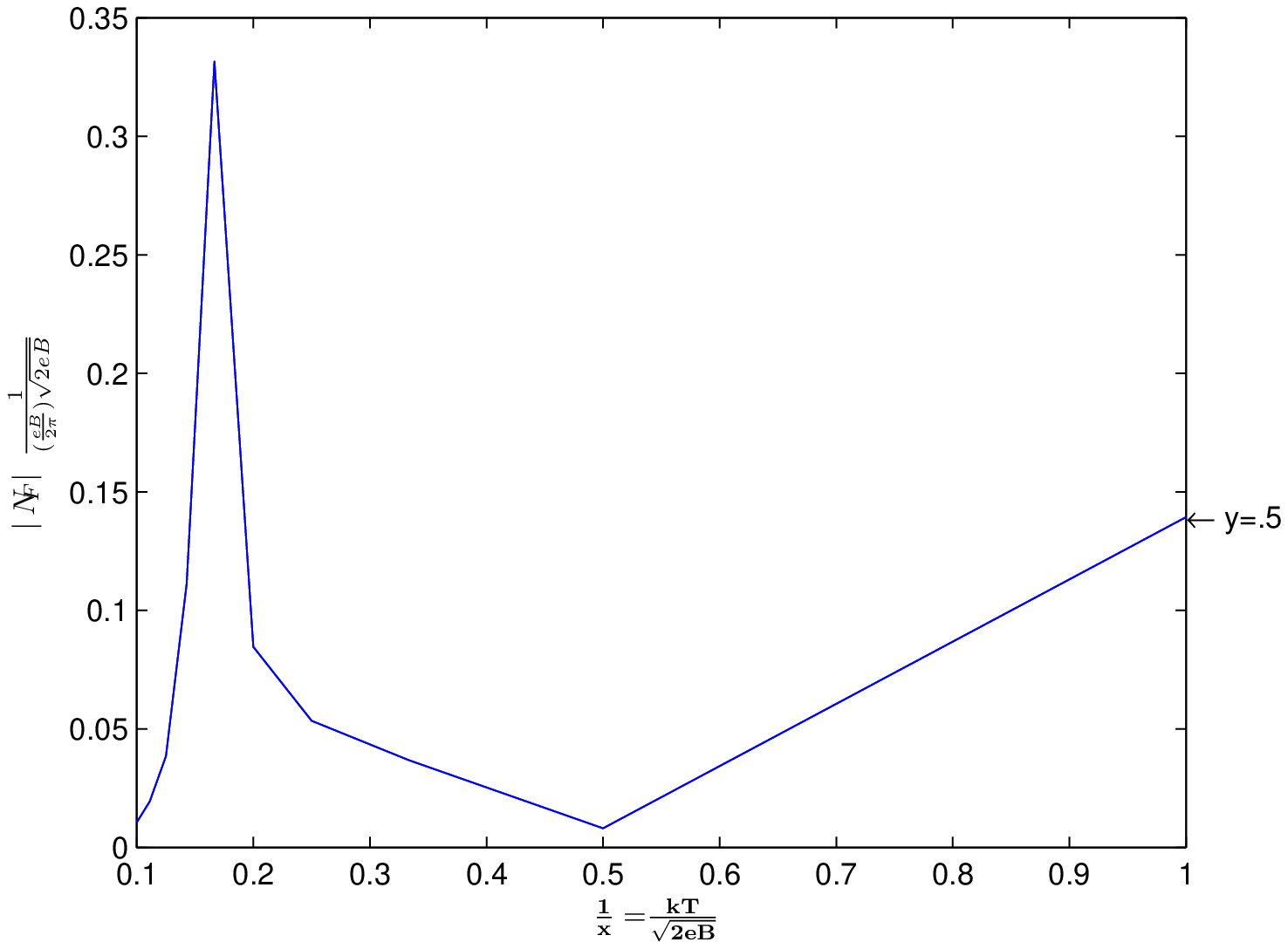}}
\caption{Variation of scaled charge density with scaled temperature for y=0.5}
\end{center}
\end{figure}

\vspace{0.5cm}

\begin{figure}
\begin{center}
\resizebox{120mm}{!}{\includegraphics{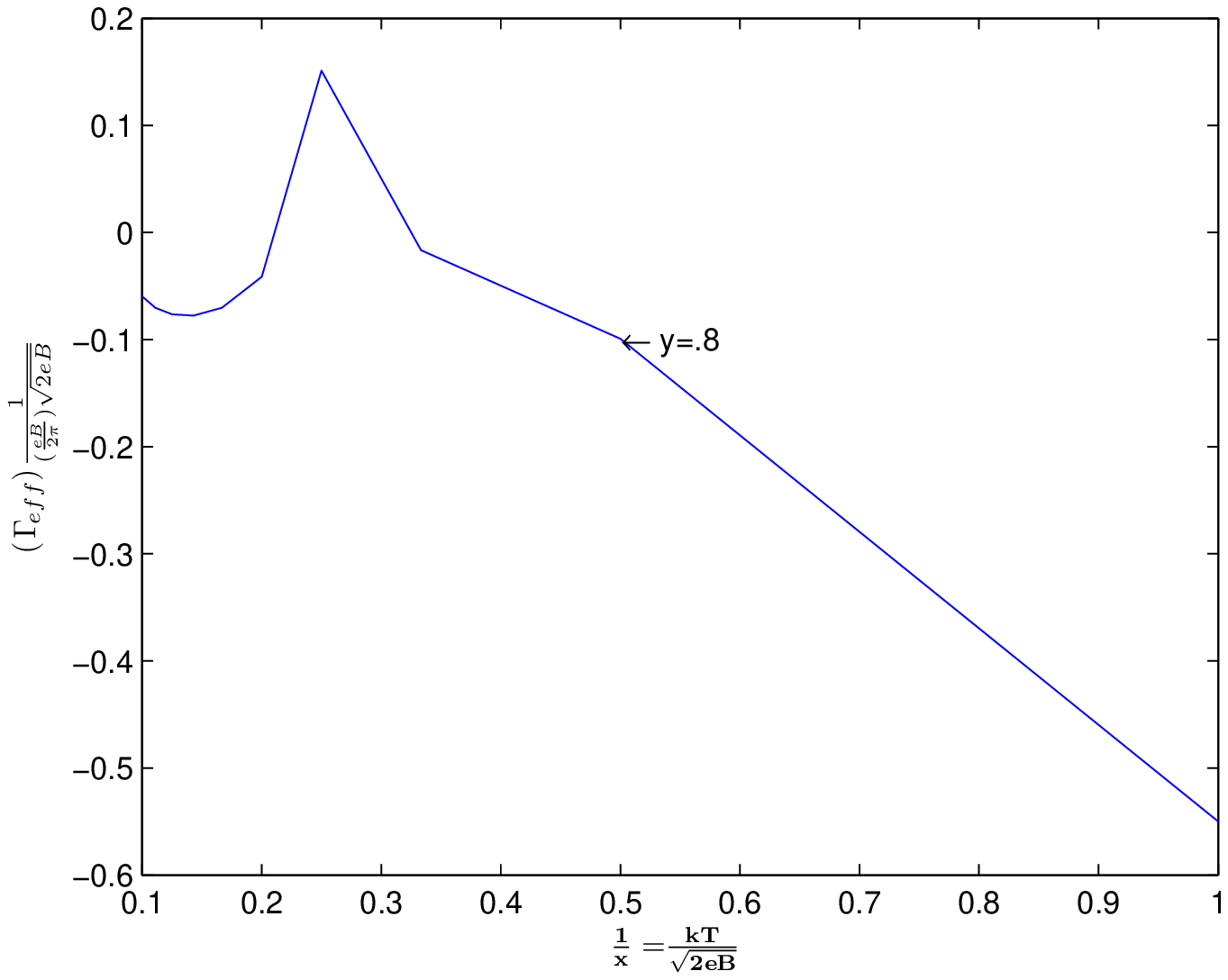}}
\caption{Variation of scaled effective energy density with scaled temperature for y=0.8}
\end{center}
\end{figure}

\vspace{0.5cm}

\begin{figure}
\begin{center}
\resizebox{120mm}{!}{\includegraphics{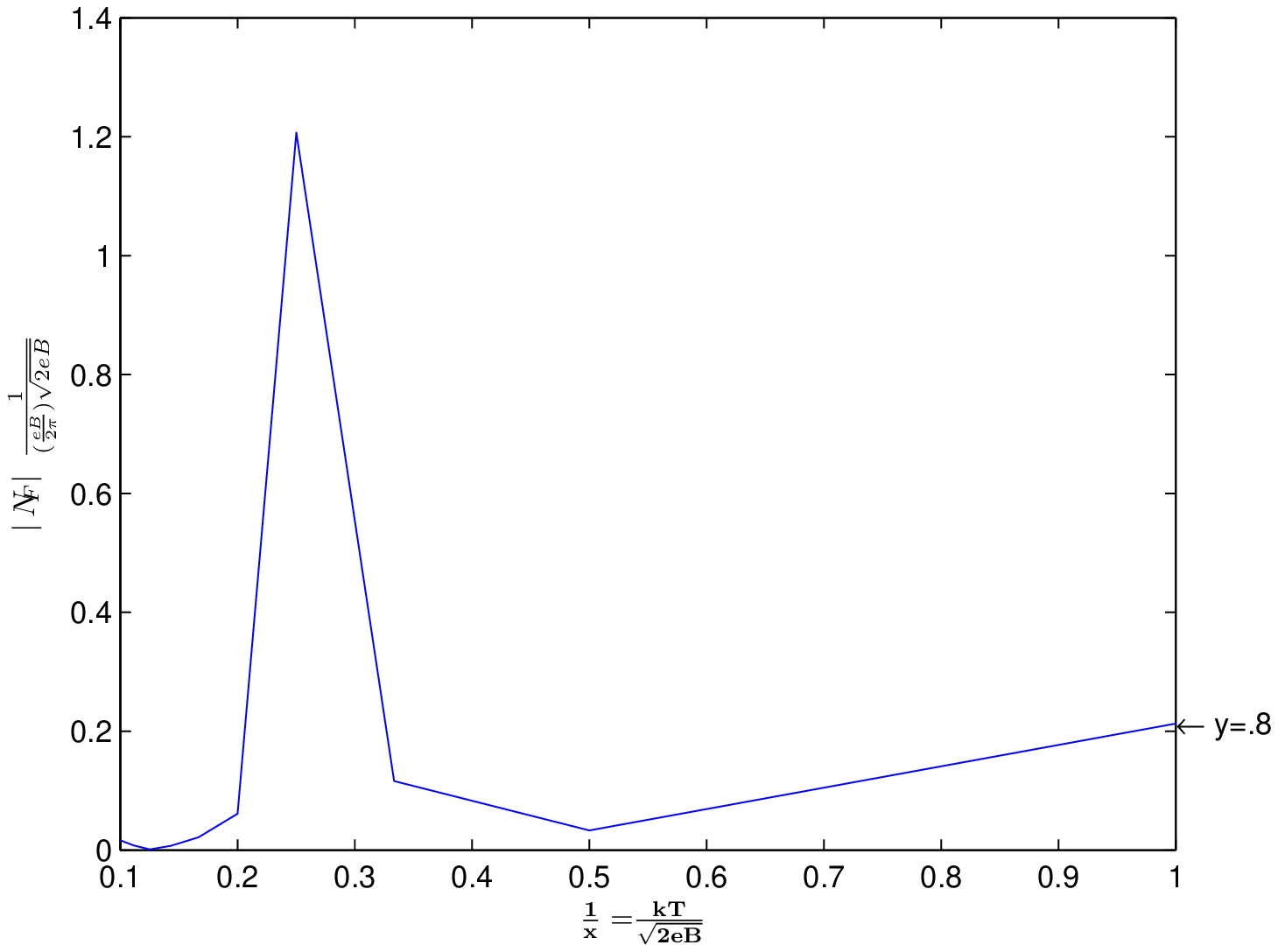}}
\caption{Variation of scaled charge density with scaled temperature for y=0.8}
\end{center}
\end{figure}
\end{document}